\documentstyle[epsfig]{espart}
\begin{document}
\begin{frontmatter}
\title{Thermal neutron induced (n,p) and (n,$\alpha$) reactions on $^{37}$Ar}
\begin{sloppypar}
\author{R.~Bieber$^{\rm a,1}$, C.~Wagemans$^{\rm b}$, G.~Goeminne$^{\rm b}$,
J.~Wagemans$^{\rm a}$,}
\author{B.~Denecke$^{\rm a}$, M.~Loiselet$^{\rm c}$, M.~Gaelens$^{\rm c}$,}
\author{P.~Geltenbort$^{\rm d}$ and H.~Oberhummer$^{\rm e}$}
\end{sloppypar}
\thanks{present address: Kernfysisch Versneller Instituut,
Rijksuniversiteit Groningen, Zernikelaan 25, NL--9744 AA Groningen, The
Netherlands}
\address{$^a$ EC, JRC, Institute for Reference Materials and Measurements,
Retieseweg, B--2440 Geel, Belgium}
\address{$^b$ Dept.~of Subatomic and Radiation Physics,
RUG,
Proeftuinstraat 86, B--9000 Gent, Belgium}
\address{$^c$ Cyclotron Research Center, UCL, Chemin
du Cyclotron 2,
B--1348 Louvain--la--Neuve, Belgium}
\address{$^d$ Institute Laue--Langevin, B.P.156, F--38042
Grenoble, France}
\address{$^e$ University of Technology Vienna, Wiedner
Hauptstr.~8--10, A--1040
Vienna, Austria}
\begin{abstract}
The $^{37}$Ar(n$_{\rm th}$,$\alpha$)$^{34}$S and
$^{37}$Ar(n$_{\rm th}$,p)$^{37}$Cl
 reactions were studied
at the high flux reactor of the ILL in Grenoble. For the
$^{37}$Ar(n$_{\rm th}$,$\alpha_0$)$^{34}$S and
$^{37}$Ar(n$_{\rm th}$,p)$^{37}$Cl
reaction cross sections,
values of
$(1070\pm80)$\,b and $(37\pm4)$\,b, respectively, were obtained.
Both values are about a factor 2 smaller than results of
older measurements.
The observed suppression of the $^{37}$Ar(n$_{\rm th}$,$\alpha_1$)$^{34}$S transition
could be verified from theoretical considerations.
Finally, evidence was found for the two--step
$^{37}$Ar(n$_{\rm th}$,$\gamma\alpha$)$^{34}$S process.
\end{abstract}
\end{frontmatter}
\section{Introduction}
One of the few nuclides with fairly large positive
reaction energies for
both proton and alpha emission after thermal neutron
capture is \nuc{37}{Ar}. 
The large
Q$_{\alpha}$--values for the transitions to
the ground and first excited state of $^{34}$S and the gap of about 2\,MeV
between
them
makes it also a good candidate for the
study of the two--step
(n$_{\rm th}$,$\gamma\alpha$) decay.

From an astrophysical point of view, the
$^{37}$Ar(n,$\alpha$)$^{34}$S and
$^{37}$Ar(n,p)$^{37}$Cl reactions occur in
nucleosynthesis networks
related to the weak s--process \cite{SCH95}. For the
calculation of the
corresponding stellar reaction rates, cross section data
are needed for neutron
energies from thermal up to a few hundred keV. In this
respect, the
reaction cross sections at thermal neutron energy are
very valuable since they
are frequently used as a normalization point for
measurements in other energy
regions and/or for the calculation of the 1/v component in
the Maxwellian averaged
cross section \cite{BAO87}. So a precise knowledge of
these cross sections is
often required.
Thermal neutron induced reaction cross
sections can also give
information about the presence of nearby resonances
\cite{BAL90}.

Neutron induced measurements on \nuc{37}{Ar} are hampered
by the fact that
this isotope is not commercially available and
furthermore has a
relatively short half life of about 35 days.
In the literature only one measurement  on the reactions
of interest
is reported by Asghar et al.~\cite{ASG78}.
It was performed more than 20 years ago and the results
have been obtained from only one
single run.
The \nuc{37}{Ar} sample used contained
1.98$\times$10$^{13}$ \nuc{37}{Ar} atoms and the
measurement resulted
in cross section values of (1970$\pm$330)\,b for the
(n$_{\rm th}$,$\alpha_0$)
and (69$\pm$14)\,b for the  (n$_{\rm th}$,p) reaction. With
such large cross
sections, the 1/v component alone can dominate the
Maxwellian averaged cross section. Therefore, the results
reported by Asghar et al.~\cite{ASG78} need to be verified and the
uncertainties
of these measurements should be reduced.

We could optimize the experimental conditions for a new measurement by
combining the
preparation of \nuc{37}{Ar} samples yielding hundred
times more
atoms than in the case of Asghar et al.~\cite{ASG78} with an
accurate determination
of the number of \nuc{37}{Ar} atoms with a dedicated
detector and a very
intense and clean thermal neutron beam.
\section{\label{sec2}Sample preparation}
One of the most crucial aspects for a precise cross
section determination on \nuc{37}{Ar}
is the preparation and the characterization of the
sample.
A detailed description of the whole procedure is given
in \cite{WLB97}.

All samples were produced at the 30\,MeV cyclotron at
Louvain--la--Neuve, Belgium.
The accelerated protons hit a NaCl target. The
\nuc{37}{Ar} atoms, created via the
$^{37}$Cl(p,n)$^{37}$Ar reaction, were ionized (2$^+$
or 1$^+$) and
afterwards implanted in a 20\,$\mu$m thin Al foil.
Samples containing up to 2$\times$10$^{15}$
\nuc{37}{Ar} atoms were produced
corresponding to about 120\,ng of \nuc{37}{Ar} in the
layer.

The number of \nuc{37}{Ar} atoms was determined at the
Institute for Reference Materials and
Measurements in Geel, Belgium, by detecting the 2.62\,keV
KX rays of
\nuc{37}{Cl},
which is created by the pure electron capture decay of
\nuc{37}{Ar}.
For this purpose a gas flow proportional detector was
used.

\begin{sloppypar}
For the present measurement four samples were used,
containing 10$^{14}$, 7.8$\times$10$^{14}$,
1.4$\times$10$^{15}$ and 2$\times$10$^{15}$ \nuc{37}{Ar} atoms,
 respectively, at the
beginning of the measurements. The uncertainty of these
values is 7\,\% and mainly due to the uncertainty
of some decay constants as explained in \cite {WLB97}.
\end{sloppypar}
\section{\label{sec3}Experimental technique}
\begin{figure}[t]
\centerline{
\epsfig{file=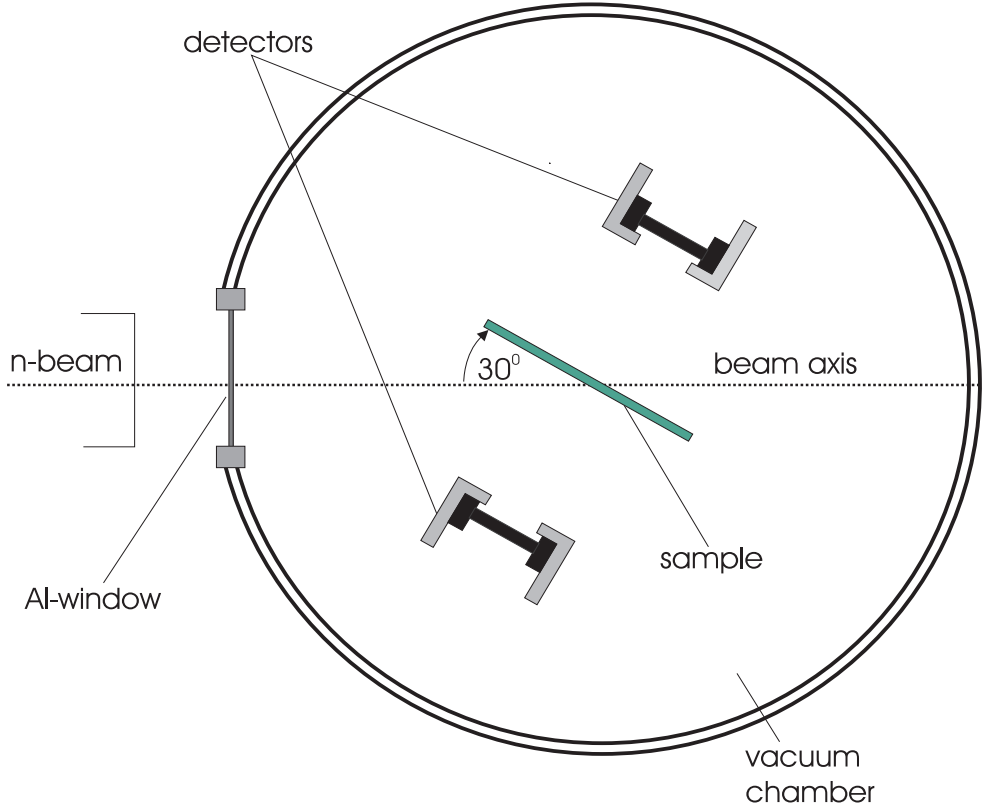,width=13cm,height=10.7cm}}
\caption{\label{fig1}Top view of the experimental
setup at the ILL.}
\end{figure}

All measurements were performed at the High Flux Reactor
at the Institute Laue--Langevin (ILL) in Grenoble, France.
The experimental setup was placed at the end of the 87\,m
curved neutron guide H22.
Due to its slight curvature, the $\gamma$--ray flux was
reduced by a factor 10$^6$.
The ratio of thermal to fast neutrons was 10$^6$.
The thermal neutron flux at the sample position was
measured to be 5$\times$10$^8$\,n/cm$^2$s.
A schematical top view of the experiment is depicted in
Fig.~\ref{fig1}.

The neutron beam entered the vacuum chamber through a thin Al window.
The sample was mounted at 30$^0$ with respect to the
neutron beam axis.
This geometry ensured that the whole sample surface was
irradiated by thermal neutrons.
Surface barrier detectors with suited thicknesses were
mounted outside the neutron beam.
Their energy calibration was done by means of the
$^{10}$B(n$_{\rm th}$,$\alpha$)$^7${Li} and
$^6$Li(n$_{\rm th}$,$\alpha$)t reactions.

All measurements were normalized via the well known
$^{235}$U(n$_{\rm th}$,f) reaction cross section,
using the value of (584.25$\pm$1.10)\,b reported in the
ENDF/B-VI data file.
This neutron flux calibration was performed strictly
maintaining the experimental geometry.
\section{\label{sec4}Measurements}
\begin{figure}[t]
\centerline{
\epsfig{file=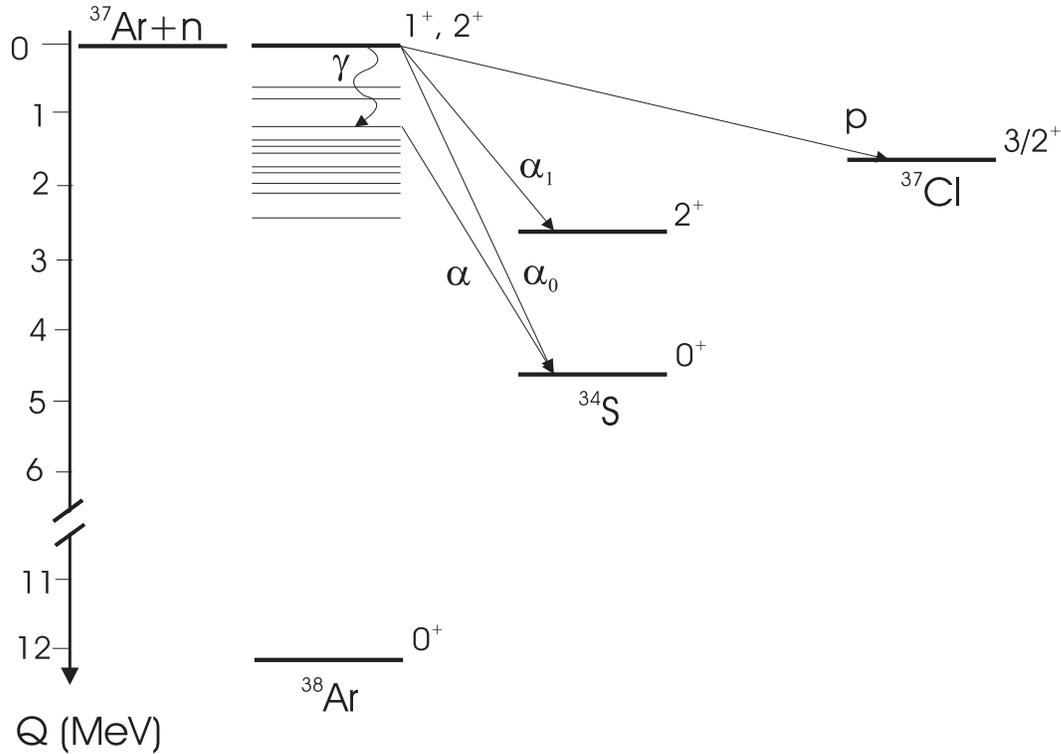,width=14cm,height=10cm}
       }
\caption{\label{fig2}Level scheme of the
$^{37}$Ar(n$_{\rm th}$,$\alpha$)$^{34}$S and
$^{37}$Ar(n$_{\rm th}$,p)$^{37}$Cl reactions.}
\end{figure}

The level scheme drawn in Fig.~\ref{fig2} illustrates
the possible reactions.
Since the ground state of \nuc{37}{Ar} is
J$^{\pi}$=3/2$^+$, only 1$^+$ and 2$^+$ levels of the
compound
nucleus \nuc{38}{Ar} will be populated during thermal
neutron capture.

For the (n$_{\mathrm {th}}$,$\alpha$)
reaction the 1$^+$ compound nucleus spin is ruled out, because
$\alpha$ emission from a 1$^+$ state to the 0$^+$
ground state of \nuc{34}{S} is parity forbidden for
initial s--waves. Most likely d--wave $\alpha$ particles emitted
from the
2$^+$ state of \nuc{38}{Ar} will determine the
(n$_{\mathrm {th}}$,$\alpha_0$) cross section.
The (n$_{\mathrm {th}}$,$\alpha_1$) transition to the
first excited state of \nuc{34}{S} having a spin/parity
assignment of 2$^+$ is
allowed, where the corresponding $\alpha$ particles
have an energy of 2.239\,MeV. Furthermore,
(n$_{\mathrm {th}}$,$\gamma\alpha$) transitions are
also possible in principle.

In the case of the (n$_{\mathrm {th}}$,p) reaction,
s--wave protons may be emitted from the 1$^+$ as well as
from the
2$^+$ state to the 3/2$^+$ ground state of
\nuc{37}{Cl}.
The Q--values and corresponding particle energies are
obtained using the masses from \cite{AUD95}:\\
${Q}_{\rm p} \hspace{0.16cm}= 1.596$\,MeV
corresponding to $E_{\rm p} \hspace{0.16cm}= 1.554$\,MeV,\\
${Q}_{\alpha_0} = 4.630$\,MeV
corresponding to $E_{\alpha_0} = 4.143$\,MeV and\\
${Q}_{\alpha_1} = 2.502$\,MeV
corresponding to $E_{\alpha_1} = 2.239$\,MeV.
\subsection{The reaction $^{37}$Ar(n$_{th}$,$\alpha$)$^{34}$S}
The choice of the detector characteristics is always a
compromise between high energy resolution (thick
detector) and low background (thin detector).
The $\alpha_0$ particles have a fairly high energy of
4.143\,MeV,
so the $^{37}$Ar(n$_{\rm th}$,$\alpha_0$)$^{34}$S reaction cross
section could be measured by using a surface
barrier detector with a thickness of 100\,$\mu$m and a
resolution of 20\,keV.
Three measuring cycles were performed with three different samples
 starting with the thinnest one.

\begin{figure}[t]
\centerline{
\epsfig{file=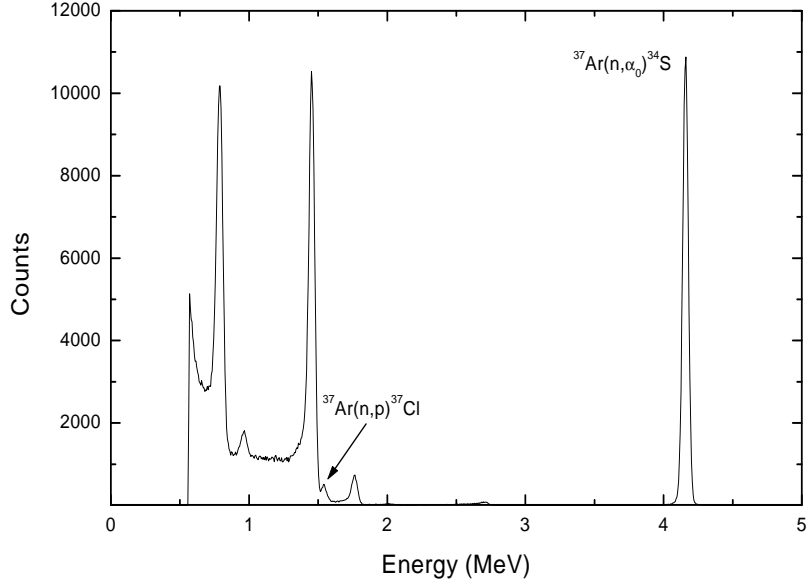,width=\textwidth,height=8cm,angle=0}
	}
\caption{\label{fig433}Typical spectrum of the
$^{37}$Ar(n$_{\rm th}$,$\alpha_0$)$^{34}$S reaction
obtained with the sample containing 1.4$\times$10$^{15}$
\nuc{37}{Ar} atoms using a 25\,$\mu$m detector.}
\end{figure}

A significant improvement concerning statistics and
signal--to--background ratio was reached
with thicker samples.
For the sample containing 1.4$\times$10$^{15}$
\nuc{37}{Ar} atoms
the $\alpha_0$ counting rate reached more than 5700
counts per hour.
In Fig.~\ref{fig433} a typical
spectrum of the $^{37}$Ar(n$_{\rm th}$,$\alpha_0$)$^{34}$S
reaction is shown using this sample.
The other peaks occurring at lower energy correspond to
the $\alpha$ and \nuc{7}{Li} lines of the $^{10}$B(n$_{\rm th}$,$\alpha$)$^{7}$Li 
reactions for \nuc{10}{B} impurities in the sample and the small peak at about
2.7\,MeV is due to the tritons produced in the
$^{6}$Li(n$_{\rm th}$,$\alpha$)t reaction.
The shoulder at about 1.55\,MeV is due to the protons
emitted in the $^{37}$Ar(n$_{\rm th}$,p)$^{37}$Cl reaction.

For the \nuc{37}{Ar}(n$_{\rm th}$,$\alpha_1$) particles,
a much lower counting rate was expected, so the
background conditions were more critical. Hence, a
25\,$\mu$m thick detector (resolution 60\,keV) was used
for this experiment. In Fig.~\ref{fig4} the result of
1100\,h of data taking is shown. Two different samples containing
respectively 5$\times$10$^{14}$ and 2$\times$10$^{15}$ \nuc{37}{Ar} atoms
were used during this measurement.
\subsection{The reaction $^{37}$Ar(n$_{th}$,$\gamma\alpha$)$^{34}$S}
\begin{figure}[t]
\centerline{
\epsfig{file=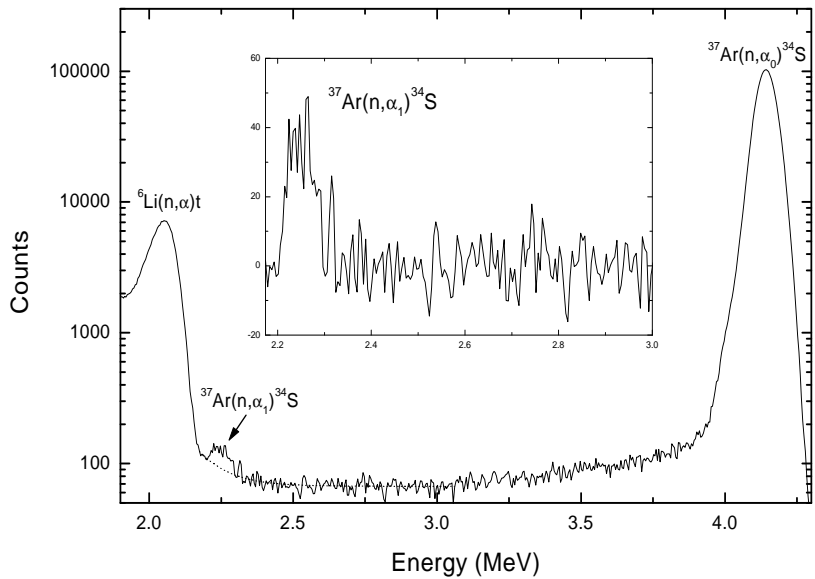,width=\textwidth,height=8cm,angle=0}
       }
\caption{\label{fig4}Spectrum of the
$^{37}$Ar(n$_{\rm th}$,$\alpha_1$)$^{34}$S and
$^{37}$Ar(n$_{\rm th}$,$\gamma\alpha$)$^{34}$S reactions
obtained with the sample containing 2$\times$10$^{15}$ $^{37}$Ar--atoms
and a 25\,$\mu$m detector. In the insert
the region of the $\alpha_1$--transition corrected with
a semi-empirical backgroundfit, is magnified.}
\end{figure}

These measurements were also performed with the sample
 containing 2$\times$10$^{15}$ \nuc{37}{Ar} atoms,
using the 25\,$\mu$m thick detector with an energy
resolution of
60\,keV. In this way, good background conditions could be
realized down to about 2\,MeV. The background
in the (n,$\gamma\alpha$) region was determined during
measurements with the reactor stopped and also with
the reactor in operation and the sample rotated over
180$^0$.
The $^{37}$Ar(n$_{\rm th}$,$\gamma\alpha$)$^{34}$S spectrum
in the region between the $\alpha_0$ and $\alpha_1$ peaks
can also be seen in Fig.~\ref{fig4}.
\subsection{The reaction$^{37}$Ar(n$_{th}$,p)$^{37}$Cl}
The problem of detecting the (n$_{\mathrm {th}}$,p)
reaction becomes evident
by looking at Fig.~\ref{fig433}. The 1.554\,MeV protons
are almost completely hidden by the
$\alpha_0$'s and $\alpha_1$'s induced in \nuc{10}{B}
impurities in the layer. Therefore, the setup used
for the (n,$\alpha$) measurements was not
suited for the determination of the (n,p) reaction.

In order to determine the
$^{37}$Ar(n$_{\rm th}$,p)$^{37}$Cl
reaction cross section, a mylar foil
(chemical composition: C$_{10}$H$_8$O$_4$, density
1.397\,g/cm$^3$) with a thickness of 5\,$\mu$m was put in
front of the detector.
Table~\ref{tab441} gives an overview of the energy loss
in
5\,$\mu$m of mylar of the main particles considered,
showing that
\begin{table}[b]
\begin{center}
\caption{\label{tab441} Overview of the energy loss of
several particles in 5\,$\mu$m mylar.}
\begin{tabular}{l|c|c|c}\hline\hline
Particle & Initial energy & Energy loss & Final energy \\
 & (MeV) & (MeV) & (MeV)\\\hline
p (\nuc{37}{Ar}) & 1.554 & 0.120 & 1.434 \\
$\alpha$ (\nuc{37}{Ar}) & 4.143 & 0.650  &  3.493\\
$\alpha_0$ (\nuc{10}{B})& 1.789 &  1.100 & 0.689 \\
$\alpha_1$ (\nuc{10}{B})& 1.483 & 1.200 & 0.283 \\
$\alpha$ (\nuc{6}{Li})  & 2.007 & 1.000 &  1.007\\
t (\nuc{6}{Li})         & 2.727 & 0.220 &
2.507\\\hline\hline
\end{tabular}
\end{center}
\end{table}
1.554\,MeV protons will only lose about 0.12\,MeV of
their energy while the
$\alpha$'s of \nuc{10}{B} are almost stopped (calculated
by using the
FORTRAN code TRIM (\underline{TR}ansport of
\underline{I}ons in
\underline{M}atter \cite {ZIE88})).
Therefore, the proton peak should emerge from the huge background
caused by \nuc{10}{B} impurities.

\begin{figure}[t]
\centerline{
\epsfig{file=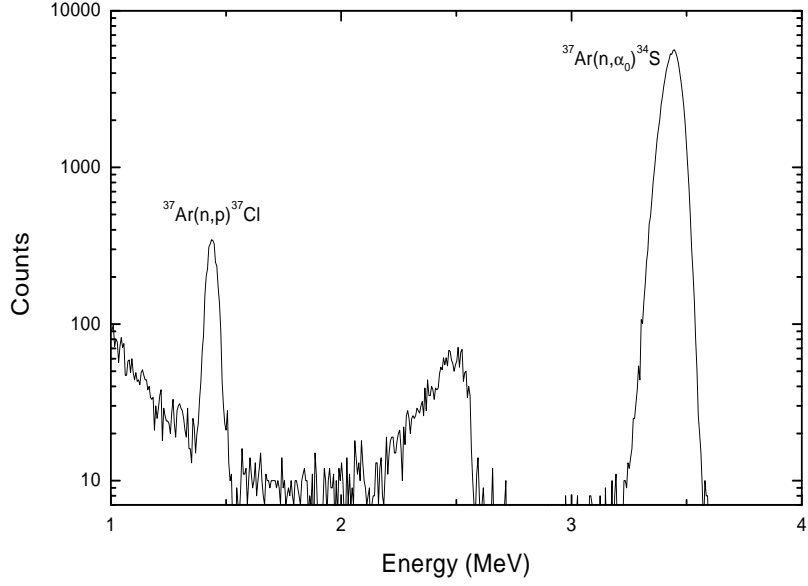,width=\textwidth,height=8cm,angle=0}
	}
\caption{\label{fig5}Spectrum of the
$^{37}$Ar(n$_{\rm th}$,p)$^{37}$Cl
reaction obtained with the sample containing
1.4$\times$10$^{15}$
\nuc{37}{Ar} atoms using a 100\,$\mu$m detector with
5\,$\mu$m of mylar in front.}
\end{figure}

The first measurement was again performed with the sample
containing 10$^{14}$ \nuc{37}{Ar}
atoms
using a 30\,$\mu$m thick detector (energy resolution of
55\,keV)
in order to have a better signal--to--background ratio.
We performed a similar measurement with a thicker
\nuc{37}{Ar} sample
(1.4$\times$10$^{15}$ \nuc{37}{Ar} atoms), this time
using a 100\,$\mu$m thick
detector with an energy resolution of 30\,keV.
Fig.~\ref{fig5} shows the re\-sult of a 16 hour measurement
under these conditions.
The count rate reached 200 protons
per hour and the background correction could easily be
done because the proton peak was situated at the end of
the low energy noise. The $^{37}$Ar(n$_{\rm th}$,p)$^{37}$Cl
cross section was determined relative to the
 particles detected in
the same measurement.
\section{\label{sec5}Results and discussion}
\subsection{The reaction $^{37}$Ar(n$_{th}$,$\alpha$)$^{34}$S}
The $^{37}$Ar(n$_{\rm th}$,$\alpha$)$^{34}$S reaction cross section
value was determined
relative to the $^{235}$U(n$_{\rm th}$,f) reaction using
following formula:
\begin{equation}
\sigma_{\alpha}=\left(\frac{N_{\rm U}}{N_{\rm Ar}}\right)
\left(\frac{C_{\alpha}}{C_{\rm f}}\right)
\left(\frac{g(T)_{\rm U}}{g(T)_{\rm Ar}}\right)\sigma_{\rm U}
\label{equ1}
\end{equation}
where N$_{\rm U}$ and N$_{\rm Ar}$ are the number of atoms/cm$^2$
of the \nuc{235}{U} and the \nuc{37}{Ar} samples,
C$_{\alpha}$ and C$_{\rm f}$ the counting rates of the
$^{37}$Ar(n$_{\rm th}$,$\alpha$)$^{34}$S and
$^{235}$U(n$_{\rm th}$,f) reactions,
g(T)$_{\rm Ar}$ and g(T)$_{\rm U}$ the corresponding Westcott
factors and $\sigma_{\rm U}$ the $^{235}$U(n$_{\rm th}$,f) reference
cross section. A detailed derivation of this formula is
given in \cite{WAG}, where also a value g(T)$_{\rm U}$ =
0.995$\pm$0.002 is reported. Since the
$^{37}$Ar(n$_{\rm th}$,$\alpha_0$)$^{34}$S cross section follows a
1/v shape below
100\,eV \cite{VOLOS}, ${\rm g(T)}_{\rm Ar}=1$ is adopted.

Due to the short half--life of \nuc{37}{Ar}, the number
of \nuc{37}{Ar} atoms present in the sample
had to be calculated for each run. The homogeneity of our
samples was checked by
mounting collimators in front of the sample.

The following (n$_{\rm th}$,$\alpha_0$) cross section
values were obtained with the three samples used:
1055\,b (10$^{14}$ at.), 1085\,b (7.8$\times$10$^{14}$
at.) and 1070\,b (1.4$\times$10$^{15}$ at.). This
illustrates the reproducibility of the measurements.
Averaging over all measurements results in a value
of
$(1070\pm80)$\,b for the
$^{37}$Ar(n$_{\rm th}$,$\alpha$)$^{34}$S reaction cross
section.
The error quoted includes all uncertainties,
the main contribution arising from the uncertainty on the
\nuc{37}{Ar} atoms in the layer.
This value is almost a factor of two lower than the one
published by Asghar et al.~\cite{ASG78}.

The $^{37}$Ar(n$_{\rm th}$,$\alpha_1$)$^{34}$S cross section was
determined relative to the
$^{37}$Ar(n$_{\rm th}$,$\alpha_0$)$^{34}$S counts recorded in the
same experiment. This yielded a
$\sigma$(n$_{\rm th}$,$\alpha_0$)/$\sigma$(n$_{\rm
th}$,$\alpha_1$) ratio of (3500$\pm$850), the error
being mainly due to the small number of events and to the
background correction. With
$\sigma$(n$_{\rm th}$,$\alpha_0$)=(1070$\pm$80)\,b this
leads to a $^{37}$Ar(n$_{\rm th}$,$\alpha_1$)$^{34}$S cross
section
of (310$\pm$100)\,mb.
This large ratio can be understood by a simple model
calculation.
Since the excitation energy of the compound nucleus
\nuc{38}{Ar} is 10.3\,MeV,
the reactions are overwhelmingly dominated by
compound nuclear
processes.
For low energies the transmission functions of the
compound nucleus are
equal to the penetration probabilities P$_{\ell}$(E),
which can be written as \cite{ROLFS}:
\begin{eqnarray}
\label{e2}
P_{\ell} (E) &  = & \exp(-2\pi\eta) \quad {\rm for} \quad
\ell=0 \\\nonumber
P_{\ell} (E) & =  & P_{0}(E)\exp\left[-2\ell(\ell+1)
\left(\frac{\hbar^2}{2\mu Z_1 Z_2 e^2 R_{\rm
n}}\right)^{\frac{1}{2}}\right]
\quad {\rm for} \quad \ell \neq 0
\end{eqnarray}
with $\eta = Z_1 Z_2 e^2/(\hbar v)$ the Sommerfeld
parameter and
$\ell$ the orbital angular momentum quantum number in the
exit channel.
$Z_1$ and $Z_2$ denote the integral charges of the
interacting nuclei
with a relative velocity $v$ and a reduced mass $\mu$.
The range of the nuclear force for the interacting nuclei
is given by $R_{\rm n}=R_0(A_1^{1/3}+A_2^{1/3})$.

\begin{sloppypar}
Since the Coulomb barrier in the exit channels of the
reaction is about
10\,MeV and thus much higher than
the corresponding energies
of the interacting
particles,
we can use Eq.~\ref{e2} to determine the penetration
probabilities in the exit channels.
The transition $^{37}$Ar(n$_{\rm th}$,$\alpha_1$)$^{34}$S
to the first excited state is suppressed
considerably more through the Coulomb barrier in the exit channel
than the $^{37}$Ar(n$_{\rm th}$,$\alpha_0$)$^{34}$S transition to the ground state,
because of the lower Q--value.
The additional hindrance through the centrifugal barrier
in the exit channel for the ground--state transition with $\ell = 2$
compared to the transition to the first excited state with $\ell = 0$
is much smaller than the dependence on the Coulomb barrier.
Based on Eq.~\ref{e2}, we calculated a cross section
ratio
$\sigma$(n$_{\rm th}$,$\alpha_0$)/$\sigma$(n$_{\rm
th}$,$\alpha_1$)\,=\,4200
that is in agreement with the experimental results.
\end{sloppypar}
\subsection{The reaction $^{37}$Ar(n$_{th}$,$\gamma\alpha$)$^{34}$S}
\begin{sloppypar}
The $^{37}$Ar(n$_{\rm th}$,$\gamma\alpha$)$^{34}$S cross section
was determined in the region between the
$^{37}$Ar(n$_{\rm th}$,$\alpha_0$)$^{34}$S and
$^{37}$Ar(n$_{\rm th}$,$\alpha_1$)$^{34}$S lines. After
a correction for the
background and the low energy tailing from the $\alpha_0$ peak,
 the counting rate in that region was
integrated and the cross section was determined relative
to
the $^{37}$Ar(n$_{\rm th}$,$\alpha_0$)$^{34}$S counting rate.
This yielded an estimation of $(9\pm3)$\,b for the $^{37}$Ar(n$_{\rm th}$,$\gamma\alpha$)$^{34}$S
cross section.
\end{sloppypar}
\subsection{The reaction $^{37}$Ar(n$_{th}$,p)$^{37}$Cl}
\begin{sloppypar}
The $^{37}$Ar(n$_{\rm th}$,p)$^{37}$Cl reaction cross section was
determined
relative to the
$^{37}$Ar(n$_{\rm th}$,$\alpha_0$)$^{34}$S
counting rate obtained in the same measurement.
The measurement with the thinnest \nuc{37}{Ar} sample
resulted in a $\sigma_{\alpha}$/$\sigma_p$ ratio of
$(27.2\pm2.6)$. The
large uncertainty quoted is due to the limited statistics
accumulated with this sample.
A great improvement was realized by using a thicker
\nuc{37}{Ar} sample. The counting rate for the
(n$_{\mathrm {th}}$,$\alpha_0$) and (n$_{\mathrm {th}}$,p) reactions was
$(92434\pm991)$ and
$(3178\pm87)$ counts, respectively, resulting
in a ratio of $(29.1\pm0.9)$. So the total uncertainty of
the ratio decreased to less than 3\,\%. These
results agree with the ratio of $(28.5\pm2.7)$ obtained
by Asghar et al.~\cite{ASG78},
which indicates that the origin of the
discrepancy between both measurements lies in the determination
 of the number of
\nuc{37}{Ar} atoms
in the sample or in the neutron flux determination.

The weighted average of our two measurements yields a
value of $(37\pm4)$\,b for the
$^{37}$Ar(n$_{\rm th}$,p)$^{37}$Cl cross section, which again is
about two times smaller than the
value of $(69\pm14)$\,b reported by Asghar et al.~.
\end{sloppypar}
\section{Conclusion}
In the present work the
$^{37}$Ar(n$_{\rm th}$,$\alpha_0$)$^{34}$S reaction cross section was
determined using three different samples.
The results were perfectly reproducible and led to an
average value of $(1070\pm80)$\,b, which is about
two times smaller than the results reported by Asghar et
al.~\cite{ASG78}. Also a weak
$^{37}$Ar(n$_{\rm th}$,$\alpha_1$)$^{34}$S transition
with a cross section of $(310\pm100)$\,mb was observed.
For the $^{37}$Ar(n$_{\rm th}$,p)$^{37}$Cl cross section we
obtained a value of $(37\pm4)$\,b.
Also $^{37}$Ar(n$_{\rm th}$,$\gamma\alpha$)$^{34}$S transitions
were clearly observed resulting in an estimation of $(9\pm3)$\,b for the cross section.


\begin{thebibliography}{9}
\bibitem{SCH95} H.~Schatz, S.~Jaag, G.~Linker,
R.~Steininger, F.~K\"appeler, P.~E.~Koehler, S.~M.~Graff
and
M.~Wiescher, {\em Phys.~Rev.\/} {\bf C51} (1995) 379.
\bibitem{BAO87} Z.~Bao and  F.~K\"appeler, {\em Atomic
Data and Nuclear Data Tables\/} {\bf 36} (1987) 411.
\bibitem{BAL90} N.~Balabanov, V.~Vtyurin, Y.~Gledenov and
J.~Popov, {\em Sov.~J.~Part.~Nucl.\/} {\bf 21} (1990) 131.
\bibitem{ASG78} M.~Asghar, A.~Emsallem, E.~Hagberg,
B.~Jonson and
P.~Tidemand--Petersson, {\em Z.~Phys.\/} {\bf A288}
(1978) 45.
\bibitem{WLB97} C.~Wagemans, M.~Loiselet, R.~Bieber,
B.~Denecke, D.~Reher and
P.~Geltenbort, {\em Nucl.~Instr. and Meth.\/} {\bf A397}
(1997) 22.
\bibitem{AUD95} G.~Audi and A.H.~Wapstra, {\em
Nucl.~Phys.\/} {\bf A595} (1995) 409.
\bibitem{ZIE88} J.F.~Ziegler and J.M.~Manoyan, {\em
Nucl.~Instr. and Meth.\/}{\bf B35} (1988) 215.
\bibitem{WAG} C.~Wagemans, P.~Schillebeeckx and
J.~P.~Bocquet, {\em Nucl.~Sci.~Eng.\/} {\bf 101} (1989)
293.
\bibitem{VOLOS}  C.~Wagemans, G.~Goeminne, R.~Bieber,
J.~Wagemans, M.~Gaelens, M.~Loiselet,
B.~Denecke, P.~Geltenbort and F.~Kolen, {\em
Proc.~Int.~Conf.~on Nuclei in the Cosmos \nolinebreak V, Volos (GR)\/}
(1998) in print.
\bibitem{ROLFS} C.~E.~Rolfs and W.~S.~Rodney, {\em
Cauldrons in the Cosmos}, The University of Chicago
Press: Chicago, 1988.
%
\end{thebibliography}
\end{document}